\documentclass[showpacs,aps,prl,twocolumn]{revtex4}
\usepackage{epsfig}
\usepackage{amsmath}

\begin{document}


\title{
{\em Ab initio} Study of the Phase Diagram of Epitaxial BaTiO$_3$
}

\author{Oswaldo Di\'eguez}
\author{Silvia Tinte}
\author{A.~Antons}
\author{Claudia Bungaro}
\author{J.~B.~Neaton} 
\altaffiliation{Present address: The Molecular Foundry, Materials Sciences
                Division, Lawrence Berkeley National Laboratory, 
                Berkeley, CA 94720, USA.}
\author{Karin M. Rabe}
\author{David Vanderbilt}

\affiliation{Department of Physics and Astronomy, Rutgers University, 
             Piscataway, New Jersey 08854-8019, USA}


\begin{abstract}
Using a combination of first-principles and effective-Hamiltonian
approaches, we map out the structure of BaTiO$_3$ under
epitaxial constraints applicable to growth on perovskite
substrates.
We obtain a phase diagram in temperature and misfit strain that is
qualitatively different from that reported by Pertsev {\em et al.}
[Phys.~Rev.~Lett. {\bf 80}, 1988 (1998)], who based their results on an
empirical thermodynamic potential with parameters fitted at temperatures in the
vicinity of the bulk phase transitions. 
In particular, we find a region of `{\em r} phase' at low temperature where
Pertsev {\em et al.} have reported an `{\em ac} phase'. 
We expect our results to be relevant to thin epitaxial films of BaTiO$_3$ at
low temperatures and experimentally-achievable strains.
\end{abstract}

\date{February 3, 2004}


\maketitle


The perovskite oxide barium titanate (BaTiO$_3$) is a prototypical
ferroelectric, an insulating solid whose macroscopic polarization can be
reoriented by the application of an electric field \cite{Lines.book.1977}.
In the perovskite ferroelectrics, it is well known both experimentally and
theoretically that the polarization is also strongly coupled to strain
\cite{Cohen.Nature.1992}, and thus that properties such as the ferroelectric
transition temperature and polarization magnitude are quite sensitive to
external stress.

Experimentally, the properties of ferroelectrics in thin film form generally
differ significantly from those in the bulk \cite{Ahn.S.2004}.
While many factors are expected to contribute to these differences, it has been
shown that the properties of perovskite thin films are strongly influenced by
the magnitude of the epitaxial strain resulting from lattice-matching the film
to the substrate.
For example, Yoneda {\em et al.}\ \cite{Yoneda.APL.1998} used molecular-beam 
epitaxy (MBE) to grow BaTiO$_3$ (lattice constant of 4.00~\AA) on
(001)-oriented SrTiO$_3$ (lattice constant of 3.91~\AA); they found that the
ferroelectric transition temperature exceeds 600~$^\circ$C, to be compared
to the bulk Curie temperature of $T_{\rm C} =$ 130~$^\circ$C.
Other studies have shown that the amount of strain in BaTiO$_3$/SrTiO$_3$
superlattices on SrTiO$_3$ substrates strongly influences properties including
the observed polarization, phase transition temperature, and dielectric
constant \cite{Chang.JAP.2000,Li.APL.2001,Shimuta.JAP.2002,Neaton.APL.2003}.

In a seminal paper, Pertsev, Zembilgotov and Tagantsev \cite{Pertsev.PRL.1998}
introduced the concept of mapping the equilibrium structure of a ferroelectric
perovskite material versus temperature and misfit strain, thus producing a
``Pertsev phase diagram'' (or Pertsev diagram) of the 
observable epitaxial phases.  
The effect of epitaxial strain is isolated from other aspects of thin-film
geometry by computing the structure of the {\em bulk} material with homogeneous
strain tensor constrained to match a given substrate with square
surface symmetry \cite{note:cubic}.
In addition, short-circuit electrical boundary conditions are imposed,
equivalent to ideal electrodes above and beneath the film
\cite{Pertsev.PRL.1998}.
Given the recognized importance of strain in determining the properties of 
thin-film ferroelectrics, Pertsev diagrams have proven to be of enormous
interest to experimentalists seeking to interpret the results of experiments
on epitaxial thin films and heterostructures.

\begin{figure}
\centerline{\epsfig{file=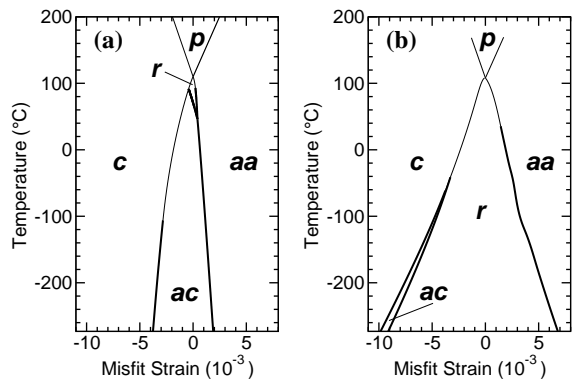,width=3.00in}}
\caption{Phase diagrams of epitaxial BaTiO$_3$ as predicted by the theory
         of Pertsev {\em et al.}\ \protect\cite{Pertsev.PRL.1998}. 
         (a) Using the parameters quoted in \protect\cite{Pertsev.PRL.1998}. 
         (b) Using the parameters quoted in \protect\cite{Koukhar.PRB.2001}.
         The second- and first-order phase transitions are represented by
         thin and thick lines, respectively.}
\label{fig:diagramsPERTSEV}
\end{figure}

In \cite{Pertsev.PRL.1998}, the mapping was carried out with a phenomenological
Landau-Devonshire model taken from the literature.
This should give excellent results in the temperature/strain regime in which
the model parameters were fitted, but will generally be less accurate when
extrapolated to other regimes.
In Fig.~\ref{fig:diagramsPERTSEV}, we compare two Pertsev diagrams for
BaTiO$_3$ computed using two different sets of Landau-Devonshire parameters,
used by Pertsev and coworkers in \cite{Pertsev.PRL.1998} and 
\cite{Koukhar.PRB.2001}, respectively.
While both give the same behavior near the bulk $T_{\rm C}$ and small misfit
strains, they predict completely different low-temperature phase behavior. 

With first-principles methods, it is possible not only to resolve such
discrepancies arising in phenomenogical theories, but also to generate a wealth
of microscopic information about the structure and properties of epitaxial
phases at various temperatures and substrate lattice constants.
In this Letter, using parameter-free total-energy methods
based on density functional theory (DFT),
we map out the equilibrium structure of BaTiO$_3$ as a function of epitaxial
constraints at zero temperature, and then extend the results to finite
temperature via an effective-Hamiltonian approach.
The Pertsev diagram obtained in this way has a similar global topology as that
of Fig.~\ref{fig:diagramsPERTSEV}(b) (but not to the one in 
\cite{Pertsev.PRL.1998}).  
This allows us to predict the impact of misfit strain on the magnitude
and orientation of the polarization and Curie temperature of BaTiO$_3$.
Our results should thus be of considerable importance for understanding 
experimental growth of high-quality, coherent epitaxial thin films of BaTiO$_3$
on perovskite substrates, as well as more generally illustrating the utility
of first-principles Pertsev diagrams.

The first-principles DFT calculations are carried out in the
Kohn-Sham framework \cite{Hohenberg.PR.1964+Kohn.PR.1965} using the
VASP software package \cite{Kresse.PRB.1993+Kresse.PRB.1996}.  The
electron-ion interaction is described by the projector augmented
wave method \cite{Blochl.PRB.1994+Kresse.PRB.1999}; semicore
electrons are included in the case of Ba ($ 5s^2 5p^6 6s^2 $) and
Ti ($ 3s^2 3p^6 4s^2 3d^2 $).  The calculations
employ the Ceperley-Alder \cite{Ceperley.PRB.1980} form of the 
local-density approximation (LDA)
exchange-correlation functional \cite{note:gga}, a 700~eV plane-wave
cutoff, and a $6 \times 6 \times 6$ Monkhorst-Pack sampling
of the Brillouin zone \cite{Monkhorst.PRB.1979}.

We begin by systematically performing optimizations of the five-atom unit
cell of the cubic perovskite structure (space group $Pm\bar{3}m$) in the
six possible phases considered by Pertsev {\em et al.}\ in 
\cite{Pertsev.PRL.1998}.
A description of these phases is given in Table \ref{table:structure}. 
Starting from a structure in which the symmetry is established by displacing
the Ti and O atoms, we relax the atomic positions and the out-of-plane
cell vector until the value of the Hellmann-Feynman forces and $zz$, $yz$ and 
$zx$ stress tensor components fall below some given thresholds
(0.001 eV/\AA~and 0.005 eV, respectively).

\begin{figure}
\centerline{\epsfig{file=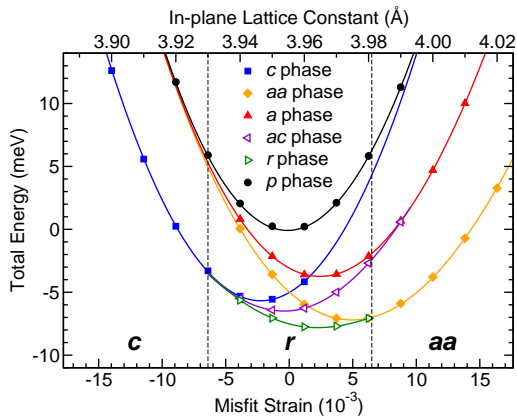,width=3.00in}}
\caption{Energies of the possible epitaxial BaTiO$_3$ phases for different
         misfit strains, as obtained from the full {\em ab initio} 
         calculations. The vertical lines denote the phase transition
         points given by the stability analysis.}
\label{fig:energiesDFT}
\end{figure}

In Fig.~\ref{fig:energiesDFT} we present the computed energy for each phase
as a function of the misfit strain $s = a/a_0 - 1$, where
$a_0$ is our DFT lattice constant for free cubic BaTiO$_3$
(3.955~\AA).
For large compressive strains, the lowest energy corresponds to the {\em c}
phase; for large tensile strains, the {\em aa} phase is favored.
At a misfit strain of $s_{\rm max}(c) = -6.4 \times 10^{-3}$ ($a = 3.930$ \AA),
there is a second-order transition
from the {\em c} phase to the {\em r} phase, with the polarization in the 
{\em r} phase continuously rotating away from the {\em z} direction as the
misfit strain increases. 
At misfit strain $s_{\rm min}(aa) = 6.5 \times 10^{-3}$ ($a = 3.981$ \AA), the
{\em r} phase polarization completes
its rotation into the {\em xy} plane, resulting in a continuous transition to the
{\em aa} phase. 
The minimum energy {\em r} phase is at misfit strain of $2.2 \times 10^{-3}$
($a = 3.964$ \AA);
lattice matching to the substrate would be optimal at this point.
At the misfit strain of the {\em c}$\rightarrow${\em r} transition, the 
polarization could also begin a continuous rotation into the (010) plane,
corresponding to the {\em ac} phase. 
However, it is clear from the figure that the energy of the {\em ac} phase is 
always higher than that of the {\em r} phase, which makes sense given that
the {\em r} phase is an epitaxial disortion of the ground-state rhombohedral
phase of bulk BaTiO$_3$, while the {\em ac} phase is related
to the higher-energy bulk orthorhombic phase.
We conclude that the phase sequence at low temperatures is not 
{\em c}$\rightarrow${\em ac}$\rightarrow${\em aa} as given in
\cite{Pertsev.PRL.1998}, but {\em c}$\rightarrow${\em r}$\rightarrow${\em aa}.

\begin{table}
\caption{Summary of possible epitaxial BaTiO$_{3}$ phases.
         In-plane cell vectors are fixed at ${\bf a}_1 = a \hat x$,
         ${\bf a}_2 = a \hat y$.  Columns list, respectively:
         phase; space group; out-of-plane lattice vector; number of
         free internal displacement coordinates; and form of the
         polarization vector.}
\begin{tabular}{ccccc}
\hline
\hline
 Phase    & SG      & ${\bf a}_3$                             & $N_{\rm{p}}$ 
                                                               & Polarization \\
\hline
 {\em p}  & $P4mmm$ & $c \hat z$                              & 0          
                                                                          & 0 \\
 {\em c}  & $P4mm$  & $c \hat z$                              & 3          
                                                                & $P_z\hat z$ \\
 {\em aa} & $Amm2$  & $c \hat z$                              & 4         
                                                         & $P(\hat x+\hat y)$ \\
 {\em a}  & $Pmm2$  & $c \hat z$                              & 4          
                                                                  & $P\hat x$ \\
 {\em ac} & $Pm$    & $c_\alpha \hat x + c \hat z$            & 8          
                                                    & $P \hat x + P_z \hat z$ \\
 {\em r}  & $Cm$    & $c_\alpha (\hat x + \hat y) + c \hat z$ & 7          
                                          & $P (\hat x+\hat y ) + P_z \hat z$ \\
\hline
\hline
\end{tabular}
\label{table:structure}
\end{table}

\begin{figure}
\centerline{\epsfig{file=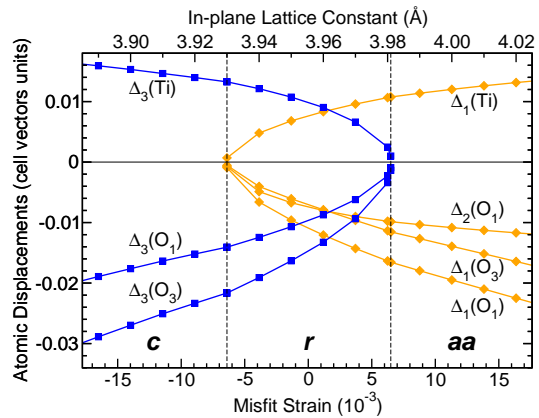,width=3.00in}}
\caption{Displacements of the atoms from the cubic perovskite cell positions,
         for the most energetically favorable configuration at a given misfit 
         strain.
         The vertical lines denote the phase transition points obtained
         from the stability analysis.
         $\Delta_3\text{(Ti)}$ labels the displacement of the Ti atom in units
         the third lattice vector, etc.}
\label{fig:displacementsDFT}
\end{figure}

Figure \ref{fig:displacementsDFT} shows the computed behavior of the atomic
displacements for the lowest-energy phase with increasing misfit strain.
For large compressive strains, the pattern of displacements corresponds to the
$c$ phase, and atoms relax only along the [001] direction.
As the in-plane strain increases, we observe a second-order phase 
transition ({\em c}$\rightarrow${\em r}), and while the magnitude of the
atomic displacements continues to diminish along [001], the displacements in
the {\em xy} plane begin to grow.
With increasing tensile strain, the displacements along [001] vanish at the
{\em r}$\rightarrow${\em aa} transition, while the displacements
in the {\em xy} plane continue to grow
smoothly.
Similar results are found when we analyze the
{\em c}$\rightarrow${\em ac}$\rightarrow${\em a} sequence (not shown), where
what was said for the {\em xy} plane applies now to the [100] direction.
The clear change in character of the displacement pattern within the
{\em r} phase witnessed here illustrates the quantitative limitations 
of using a single misfit-strain-independent local mode to model the 
phase diagram. 

A stability analysis provides the precise
limits of phase stability shown in Figs.~\ref{fig:energiesDFT}
and \ref{fig:displacementsDFT}.  At each value of misfit in the {\em c}
phase, for example, we carry out finite-difference calculations of
$x$ forces and $xz$ stress as the atomic $x$ coordinates and
$xz$ strain are varied.  The zero crossing of the lowest eigenvalue
of the resulting $6 \times 6$ Hessian matrix identifies
the critical misfit.  A similar analysis is used to consider $z$
displacements and shear strains in the {\em a} and {\em aa} phases.
By properly considering zone-center phonons, elastic
shear, and linear cross-coupling between them, this analysis allows
us to locate the second-order phase boundaries much more precisely
than is possible through direct comparison of total energies \cite{note:ifc}.

Having established the first principles zero-temperature phase diagram,
we now extend our study of epitaxial BaTiO$_3$ to finite temperatures using
the effective Hamiltonian approach of Zhong, Vanderbilt, and
Rabe~\cite{Zhong.PRL.1994+Zhong.PRB.1995}. 
In this method, the full Hamiltonian is mapped onto a
statistical mechanical model by a subspace projection, and parameterized
through {\em ab initio} calculations of small distortions of bulk BaTiO$_3$ in
the cubic perovskite structure.
The reduced subspace is composed of a set of relevant degrees of freedom
identified for ferroelectric perovskites as the unit cell distortions
corresponding to local polarization, expressed in the form of local modes.
This subspace is augmented by the inclusion of the homogeneous strain.

\begin{figure}
\centerline{\epsfig{file=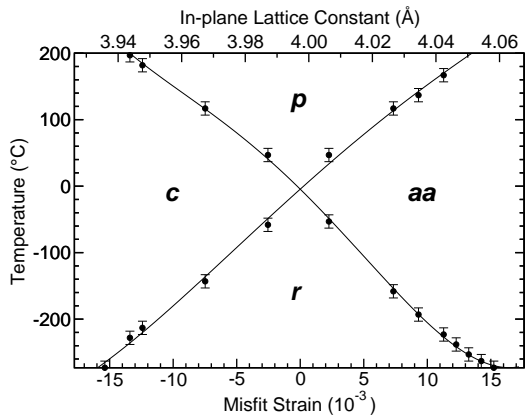,width=3.00in}}
\caption{Phase diagram of epitaxial BaTiO$_3$ obtained using the effective 
         Hamiltonian of Zhong, Vanderbilt and Rabe
         \protect\cite{Zhong.PRL.1994+Zhong.PRB.1995}.}
\label{fig:diagramHEFF}
\end{figure}

It is straightforward to impose the constraint of fixed in-plane strain by
fixing three of the six tensor strain components during the Monte Carlo
(MC) simulations.  For each value of in-plane strain, MC
thermal averages are obtained for the unconstrained components of the
homogeneous strain and the average polarization \cite{note:mcdetails},
and phase transitions are identified by monitoring the symmetry of these
quantities.
Following \cite{Zhong.PRL.1994+Zhong.PRB.1995}, all the simulations were
performed at the same negative external pressure of $P=-4.8$~GPa.
Misfit is defined relative to $a_0=3.998$~\AA, the lattice
constant at the bulk cubic-to-tetragonal transition as computed
with this approach \cite{Zhong.PRL.1994+Zhong.PRB.1995}.
The resulting phase diagram appears in Fig.~\ref{fig:diagramHEFF}, where all
phase lines represent second-order transitions.

The Pertsev diagrams of Figs.~\ref{fig:diagramsPERTSEV}(a),
\ref{fig:diagramsPERTSEV}(b), and \ref{fig:diagramHEFF} share the
same topology above and just below $T_{\rm C}$: {\em p} at high
temperature, {\em c} at large compressive misfit, {\em aa} at 
large tensile misfit, and
a 4-phase point connecting these phases with the {\em r} phase at
$T_{\rm C}$.  At lower temperature, there is a drastic difference
between Figs.~\ref{fig:diagramsPERTSEV}(a) and
\ref{fig:diagramsPERTSEV}(b), with our theory supporting the
latter.  While our theory underestimates the temperature of
the 4-phase crossing point in Fig.~\ref{fig:diagramHEFF} by about
100~$^\circ$C, this is the price we pay for insisting on a
first-principles approach; indeed, this effective Hamiltonian
underestimates the temperature of the bulk cubic-to-tetragonal
transition by about the same amount.

At low temperature, our Pertsev diagram shows the sequence of second-order phase
transitions {\em c}$\rightarrow${\em r}$\rightarrow${\em aa}.
The {\em r} phase is predicted to exist in a range that is more than twice
as broad as that shown in Fig.~\ref{fig:energiesDFT}.
We have found that this range is reduced to about 1.5 times that of Fig.~2 
when the negative-pressure correction is not included.
The remaining discrepancy is related to technical differences between the DFT 
calculations used in \cite{King-Smith.PRB.1994} to obtain the parameters for 
the effective Hamiltonian method and the DFT calculations we report here
\cite{note:technical}.
We should also mention that the effective Hamiltonian used here does not
include the physics related to the zero-point motion of the ions.
This quantum effect should alter the shape of the lines of the diagram at very 
low temperatures, and it would result in those lines approaching the
misfit-strain axis with infinite derivative (see, for example,
\cite{Iniguez.PRL.2002}).
In any case, at zero temperature, the phase sequence is quite unambiguously
established by the first-principles results.
This clearly indicates that the low-temperature extrapolation of the
Landau-Devonshire parameters fitted near $T_{\rm C}$ can give rise to spurious 
results, such as the stability of the {\em ac} phase obtained in
Ref. \cite{Pertsev.PRL.1998}. 

Finally, we comment on the effect of the assumptions made in the construction
of this first-principles Pertsev diagram. 
In principle, we should consider the possibility of equilibrium structures with
larger unit cells, particularly those with cell-doubling octahedral rotations,
which have been shown to be important in SrTiO$_3$, and could condense in
BaTiO$_3$ under sufficiently large misfit strains.
As an example, we have checked that the paraelectric phase of the film is
stable with respect to octahedral rotations about the [001] direction (with
M$_3$ symmetry) up to an epitaxial compressive strain of 
$-70.9 \times 10^{-3}$ ($a = 3.675$ \AA), far larger than those likely to be 
experimentally relevant.
In addition, while we have studied only the effects of epitaxial strain,
other physical effects may also be relevant to the structure and properties of
thin films, such as atomic rearrangements at the film-substrate interface and
free surface, and the instability to formation of multiple domain
structures \cite{Li.APL.2003}.

To summarize, we have performed density-functional theory calculations in order
to obtain the ``Pertsev diagram'' of epitaxial BaTiO$_3$ at zero temperature.
The results we obtain differ from those computed previously 
\cite{Pertsev.PRL.1998} using a Landau-Devonshire theory where the 
parameters needed were obtained from experimental information about bulk
BaTiO$_3$ at the phase transitions temperatures.
Alternatively, the use of a similar theory where the constants of the model
are computed using an {\em ab initio} method is consistent with both the 
first principles results at zero temperature, and with the work of Pertsev 
{\em et al.} \cite{Pertsev.PRL.1998} at high temperature.

It is a pleasure to thank 
Jorge \'I\~niguez, 
Javier Junquera,
and Jos\'e Juan Blanco-Pillado
for useful discussions. 
This work was supported by ONR Grants N0014-97-1-0048, N00014-00-1-0261, and
N00014-01-1-0365, and DOE Grant DE-FG02-01ER45937.




\begin{thebibliography}{0}

\bibitem{Lines.book.1977}
M.~E.~Lines and A.~M.~Glass, 
{\em Principles and Applications of Ferroelectrics and Related Materials}
(Clarendon Press, Oxford, 1977). 

\bibitem{Cohen.Nature.1992}
R.~E.~Cohen,
Nature {\bf 358}, 136 (1992).

\bibitem{Ahn.S.2004}
C.~H.~Ahn, K.~M.~Rabe, J.-M.~Triscone, 
Science {\bf 303}, 488 (2004).

\bibitem{Yoneda.APL.1998}
Y.~Yoneda, T.~Okabe, K.~Sakaue, H.~Terauchi, H.~Kasatani, and K.~Deguchi,
J. Appl. Phys. {\bf 83}, 2458 (1998).

\bibitem{Chang.JAP.2000}
W.~Chang, C.~M.~Gilmore, W.-J.~Kim, J.~M.~Pond, S.~W.~Kirchoefer, S.~B.~Qadri,
D.~B.~Chrisey, and J.~S.~Horwitz,
J. Appl. Phys. {\bf 87}, 3044 (2000).

\bibitem{Li.APL.2001}
H.~Li, A.~L.~Roytburd, S.~P.~Alpay, T.~D.~Tran, L.~Salamanca-Riba, and
R.~Ramesh,
Appl. Phys. Lett. {\bf 78}, 2354 (2001).

\bibitem{Shimuta.JAP.2002}
T.~Shimuta, O.~Nakagawara, T.~Makino, S.~Arai, H.~Tabata, and T.~Kawai,
J. Appl. Phys. {\bf 91}, 2290 (2002).

\bibitem{Neaton.APL.2003}
J.~B.~Neaton and K.~M.~Rabe, Appl. Phys. Lett. 82, 1586 (2003).

\bibitem{Pertsev.PRL.1998}
N.~A.~Pertsev, A.~G.~Zembilgotov, and A.~K.~Tagantsev,
Phys. Rev. Lett. {\bf 80}, 1988 (1998).

\bibitem{note:cubic}
The applicability of our study is thus not limited to cubic substrates; it
also applies, e.g., to tetragonal perovskite substrates.

\bibitem{Koukhar.PRB.2001}
V.~G.~Koukhar, N.~A.~Pertsev, and R.~Waser,
Phys. Rev. B {\bf 64}, 214103 (2001).

\bibitem{Hohenberg.PR.1964+Kohn.PR.1965}
P.~Hohenberg and W.~Kohn,
Phys. Rev. {\bf 136}, B864 (1964); 
W.~Kohn and L.~J.~Sham,
Phys. Rev. {\bf 140}, A1133 (1965).

\bibitem{Kresse.PRB.1993+Kresse.PRB.1996}
G.~Kresse and J.~Hafner, 
Phys. Rev. B {\bf 47}, 558 (1993);
G.~Kresse and J.~Furthm\"uller,
Phys. Rev. B {\bf 54}, 11169 (1996).

\bibitem{Blochl.PRB.1994+Kresse.PRB.1999}
P.~E.~Bl\"ochl,
Phys. Rev. B {\bf 50}, 17953 (1994); 
G.~Kresse and D.~Joubert,
Phys. Rev. B {\bf 59}, 1758 (1999).

\bibitem{Ceperley.PRB.1980}
D.~M.~Ceperley and B.~J.~Alder,
Phys. Rev. Lett. {\bf 45}, 566 (1980).

\bibitem{note:gga}
Using the generalized gradient approximation (GGA) instead of the LDA does
not lead to substantial improvements in the case of BaTiO$_3$; see 
D. J. Singh,
Ferroelectrics {\bf 164}, 143 (1995).

\bibitem{Monkhorst.PRB.1979}
H.~J.~Monkhorst and J.~D.~Pack,
Phys. Rev. B {\bf 13}, 5188 (1976). 

\bibitem{note:ifc}
Similarly, the {\em a} phase is stable against the {\em ac} phase down
to $s_{\rm min}(a) = 10.3 \times 10^{-3}$ ($a = 3.993$ \AA).
Omitting the strain and working only with the
$5 \times 5$ Hessian matrix matrix results in very little error:
$s_{\rm max}(c) = -6.2 \times 10^{-3}$ ($a = 3.931$ \AA), 
$s_{\rm min}(aa) = 6.1 \times 10^{-3}$ ($a = 3.980$ \AA), and
$s_{\rm min}(a) =  9.3 \times 10^{-3}$ ($a = 3.992$ \AA).

\bibitem{Zhong.PRL.1994+Zhong.PRB.1995}
W.~Zhong, D.~Vanderbilt, and K.~M.~Rabe,
Phys. Rev. Lett. {\bf 73}, 1861 (1994);
W.~Zhong, D.~Vanderbilt, and K.~M.~Rabe,
Phys. Rev. B {\bf 52}, 6301 (1995).

\bibitem{note:mcdetails}
MC simulations were performed using a 12$\times$12$\times$12 supercell.
Typically 30,000~MC sweeps were used to equilibrate the system, and
an additional 50,000 to obtain averages of local-mode variables with a
statistical error below $10\%$.  The temperature was increased in steps
of 5~K.

\bibitem{King-Smith.PRB.1994}
R.~D.~King-Smith and D.~Vanderbilt, 
Phys. Rev. B {\bf 49}, 5828 (1994).

\bibitem{note:technical}
In particular, the soft-mode eigenvalue has been found to be very sensitive to
the fineness of the grid used to evaluate the Fourier transforms.
This discrepancy vanishes almost completely if instead of working with the
soft-mode eigenvalue calculated in \protect\cite{King-Smith.PRB.1994}
(0.0350 a.u.) we use the one given by our new DFT calculations
(0.0223 a.u.) that was calculated using a finer grid than
in \protect\cite{King-Smith.PRB.1994}.

\bibitem{Iniguez.PRL.2002}
J.~\'I\~niguez and D.~Vanderbilt,
Phys. Rev. Lett. {\bf 89}, 115503 (2002).

\bibitem{Li.APL.2003}
Y.~L.~Li, S.~Choudhury, Z.~K.~Liu, and L.~Q.~Chen, 
Appl. Phys. Lett. {\bf 83}, 1608 (2003). 

\end{thebibliography}
\end{document}